\documentclass[pra,twocolumn,superscriptaddress,10pt,noshowpacs]{revtex4}
\usepackage[english]{babel}
\usepackage[T1]{fontenc}
\usepackage[utf8]{inputenc}
\usepackage{graphicx,epstopdf}
\usepackage{amsmath}
\DeclareMathOperator{\csch}{csch}
\usepackage{amsfonts}
\usepackage{bbm}
\usepackage{amssymb}
\usepackage{color}
\usepackage{latexsym}
\usepackage{caption}
\usepackage{subcaption}
\usepackage{times,txfonts}

\DeclareMathOperator{\sech}{sech}
\begin{document}

\title{Electronic states in a quantum Beltrami surface}

\author{J. Furtado\footnote{E-mail: job.furtado@ufca.edu.br}}

%\affiliation{Department of Physics, Faculty of Science, Gazi University, 06500 Ankara, Turkey}
\affiliation{Centro de Ci\^{e}ncias e Tecnologias, Universidade Federal do Cariri, 63048-080, Juazeiro do Norte, Brazil}

\date{\today}

%ABSTRACT----------------------------------------------------------------

\begin{abstract}

In this paper, we investigate the influence of the geometry in the electronic states of a quantum Beltrami surface. We have considered an electron governed by the spinless stationary Schr\"{o}dinger equation constrained to move on the Beltrami surface due to a confining potential from which the Da Costa potential emerges. We investigate the role played by the geometry and orbital angular momentum on the electronic states of the system.  

\end{abstract}

\maketitle

%INTRODUCTION-------------------------------------------------------------

\section{Introduction}

Quantum mechanics at curved surfaces brings about fascinating phenomena. Quantum particles exhibit wave-like properties, such as diffraction and interference, even on curved surfaces. When dealing with curved surfaces, such as the surface of a torus \cite{GomesSilva:2020fxo}, catenoid \cite{euclides}, among others, the principles of quantum mechanics take on a fascinating twist. The curvature of the surface introduces a geometric factor that significantly influences the behavior of quantum particles. These phenomena give rise to intriguing effects, such as the formation of quantized energy levels and the emergence of geometric phases \cite{costa, costa1}. Additionally, the curvature of the surface can induce curvature-induced forces on the particles, influencing their trajectories and dynamics \cite{costa, costa1}. Exploring the interplay between quantum mechanics and curved surfaces unveils a rich landscape of phenomena, providing insights into fundamental aspects of both quantum theory and geometry.

Important two-dimensional nanostructures in low energy physics, such as graphene \cite{katsnelson, geim, castro} and phosphorene \cite{phosphorene}, have attracted attention due to their unusual properties. The elec tronic properties of such two-dimensional systems are highly dependent on the geometry \cite{CostaFilho:2020sbw, Aguiar:2020dgi}, so that they can be used as analog models for high energy physics systems \cite{Capozziello:2020ncr, Cvetic:2012vg, Pourhassan:2018wjg, Acquaviva:2022yiq, Iorio:2013ifa, Iorio:2010pv, Iorio:2014pwa}.

%Parágrafo adaptado do artigo com a Ozlem (torus)
Among the studied surfaces we may highlight the torus. On the toroidal geometry the curvature effects plays a very significant role \cite{GomesSilva:2020fxo, encinosa}. Carbon nanotori are present in nanoelectronics, biosensors and quantum computing, \cite{Goldsmith, Goldsmith2}. Considering an electron with dynamic governed by the Schr\"{o}dinger equation, the curvature-induced bound-state eigenvalues and eigenfunctions were calculated in \cite{encinosa} for a particle constrained to move on a toroidal surface. These same considerations were considered and the action of external fields was investigated in \cite{GomesSilva:2020fxo}. Charged spin 1/2 particle, governed by Pauli equation, moving along a surface of a torus was studied in \cite{AGM}. Analytical solutions for the (2 + 1) Dirac equation on the torus surface were first obtained in \cite{Yesiltas:2018zoy} using a supersymmetric approach of quantum mechanics for two cases, constant and position-dependent Fermi velocity. These same considerations were taken into account for the investigation of the Dirac equation on the torus under the action of external fields in \cite{Yesiltas:2021crm}. More recently, a possible qubit encoding in the energy levels of a graphene nanotorus was investigated in \cite{Furtado:2022uvk}.

%Parágrafo adaptado do artigo com a Ozlem (catenoid)
The catenoid geometry is drawing a lot of attention in the last years. As a minimal surface, some two dimensional wormhole geometries are equivalent to catenoid shaped surfaces \cite{dandoloff}. In Ref. \cite{gonzalez, pincak} the authors proposed a bridge connecting a bilayer graphene using a nanotube. In order to obtain a smooth bridge, Ref.\cite{dandoloff, dandoloff2}, suggested a catenoid surface aiming to describe the bilayer and the bridge using a single surface. The catenoid curvature is concentrated around the bridge and vanishes asymptotically \cite{spivak}. For non-relativistic electrons, the surface curvature gives rise to a geometric potential in the Schr\"{o}dinger equation. The effects of the external electric and magnetic fields and geometry upon the graphene catenoid bridge was investigated in Ref.\cite{euclides}, where it was considered a single electron whose dynamics is governed by the Schr\"{o}dinger equation on the surface. Also, the consideration of a position-dependent mass problem upon the electron on a catenoid bridge was addressed in Ref. \cite{Yesiltas:2021dpm}, where it was proposed an isotropic position-dependent mass as a function of the Gaussian and mean curvatures. And more recently, the electronic states of the generalized Ellis-Bronnikov bilayer graphene wormhole were studied \cite{deSouza:2022ioq}. 

As far as we know, only a few works were carried out regarding the study of the properties of a quantum Beltrami surface, see f.e. \cite{Morresi, Gallerati}. In \cite{Morresi}, the authors investigate Event Horizons and Hawking radiation through a Beltrami shaped graphene membrane. In \cite{Gallerati} the author study the electronic properties of Beltrami shaped graphene-like material where analytical solutions were obtained for charge carriers described by the massless Dirac equation. The Beltrami surface is a special type of surface that possesses constant Gaussian curvature and exhibits interesting geometric properties. Hence, in this paper, we investigate the influence of the geometry in the electronic states of a quantum Beltrami surface. We have considered an electron governed by the spinless stationary Schr\"{o}dinger equation constrained to move on the surface of a pseudosphere due to a confining potential from which the Da Costa potential emerges. We investigate the role played by the geometry and orbital angular momentum on the electronic states of the system.  

This paper is organized as follows: In section II we present the dynamics of an electron governed by the Schr\"{o}dinger equation constrained to move on a Beltrami surface. In section III we calculate the bound states for the electron on the Beltrami surface considering both the geometric and centrifugal effects. In section IV we draw our conclusions.  

%Pseudosphere surface----------------------------

\section{Electron on a Beltrami surface}
\label{section2}

\begin{figure*}[ht!]
    \centering
\begin{subfigure}{0.4\textwidth}
  \centering
  \includegraphics[scale=0.3]{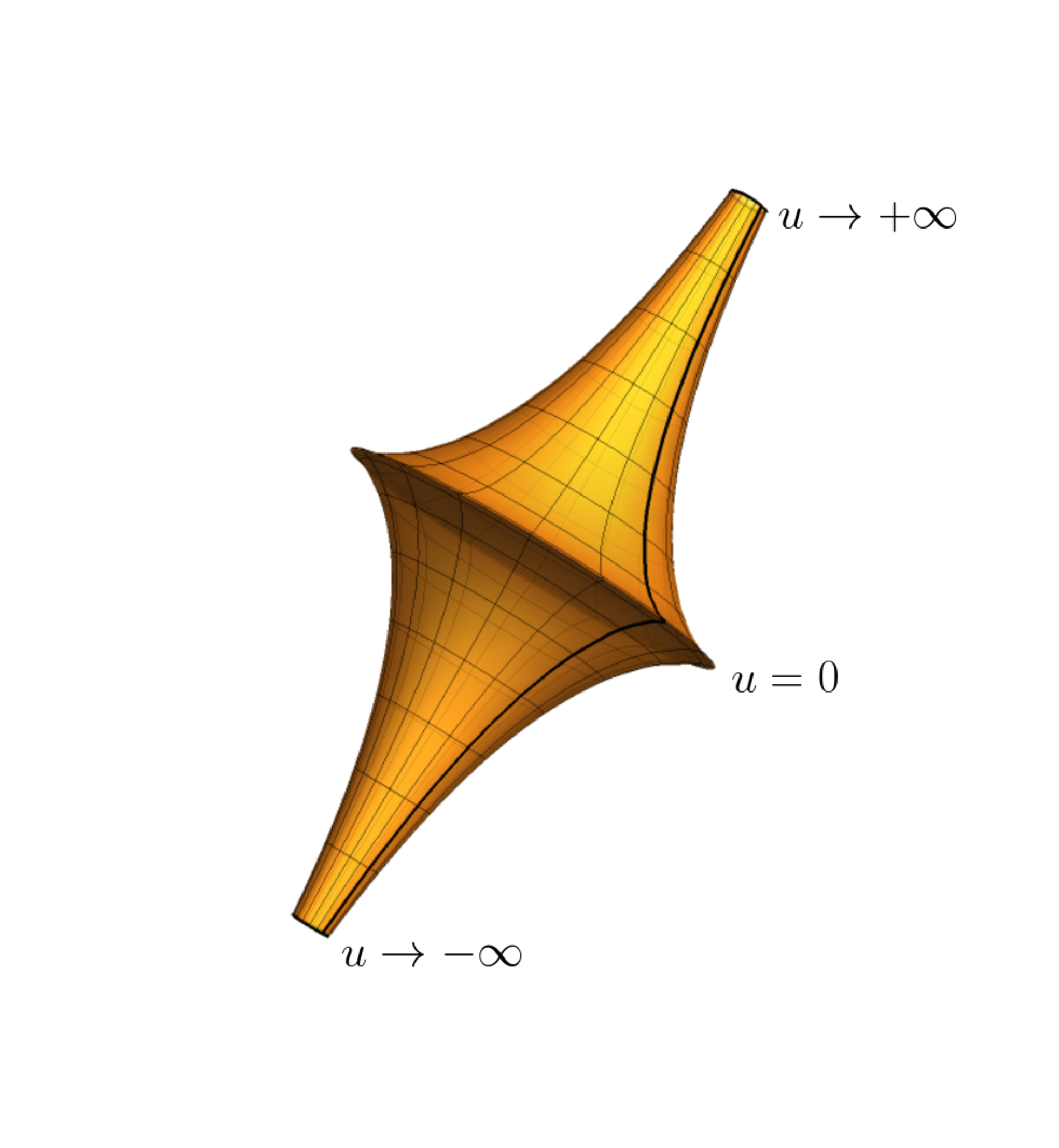}
  \caption{}
 
\end{subfigure}%
\begin{subfigure}{.6\textwidth}
  \centering
  \includegraphics[scale=0.45]{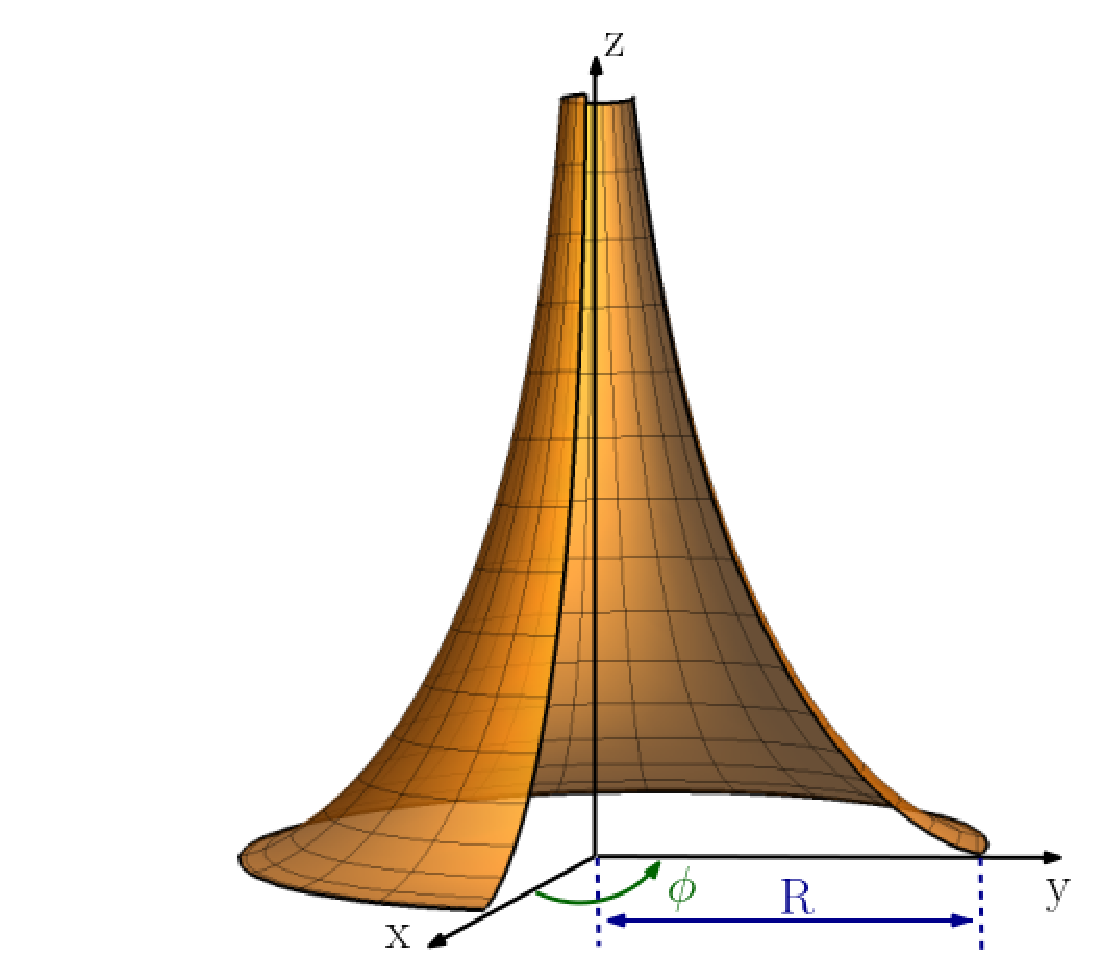}
  \caption{}

\end{subfigure}%
    \caption{Beltrami surface and coordinate system.}
    \label{Fig1}
    
\end{figure*}

In this section, we introduce the geometry and the dynamics of the electron on the Beltrami surface considering the effects of the curvature and geometry of the system. We will be considering an electron described by a quadratic dispersion relation, so that we employee the Schr\"{o}dinger equation in curved surfaces in order to properly describe the system. As shown in fig.\ref{Fig1}, the Beltrami surface is realized as the revolution surface generated by the rotation of the tractrix curve about its asymptote. Our coordinate system is depicted in fig.\ref{Fig1}.

After squeezing the electron wave-function on the surface, the spinless stationary Schr\"{o}dinger equation has the form
\begin{equation}
\label{constantmassschrodinger}
    -\frac{\hbar^2}{2m^*}\nabla^2 \Psi +V_{dc}\Psi=E\Psi,
\end{equation}
where $\nabla^{2}\Psi=\frac{1}{\sqrt{g}}\partial_{a}(\sqrt{g}g^{ab}\partial_{b}\Psi)$ is the Laplacian operator on the surface, $g^{ab}$ is the induced metric of the surface, $m^*$ is the effective mass of the electron and $V_{dC} = - \frac{\hbar^2}{2m^{*}}(H^2-K)$ is a potential induced by the surface curvature, known as the da Costa potential \cite{costa}. The geometric potential depends both on the  the mean curvature $H$ and the Gaussian curvature $K$ \cite{costa}. 

We adopt the Beltrami coordinates as
\begin{eqnarray}
\label{cilindricalcoordinates}
\nonumber \vec{r}(u,\phi)&=&R\sech\left(\frac{u}{R}\right)\cos(\phi)\hat{i} + R\sech\left(\frac{u}{R}\right)\sin(\phi)\hat{j} +\\
&&+ \left(u-R\tanh\left(\frac{u}{R}\right)\right)\hat{k},
\end{eqnarray}
where $R$ is the radius of the pseudosphere in $u=0$. The coordinates $u\in(-\infty,\infty)$ and $\phi\in[0,2\pi)$ cover the whole pseudosphere. It is important to highlight that for a real two-dimensional material, some cutoffs in the $u$-coordinate must be imposed in order to properly guarantee the validity of the continuum limit employed here. For a graphene-like material, for example, the bonds in the graphene lattice are around $1,43 \AA$ \cite{graphene-bond}, so that the radius $r$ for a given value of $u$ must be $r>>1,43 \AA$ in order to ensure the continuum limit approach.

\begin{figure}[h!]
    \centering
    \includegraphics[scale=0.8]{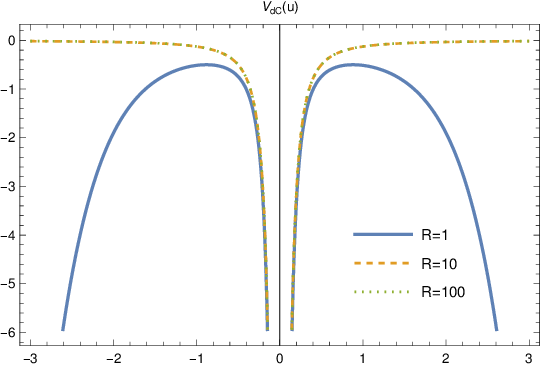}
    \caption{Da Costa potential for the pseudosphere.}
    \label{dacosta}
\end{figure}

In this coordinate system, the interval reads
\begin{eqnarray}
ds^{2}=\tanh ^2\left(\frac{u}{R}\right)du^2+R^2 \sech^2\left(\frac{u}{R}\right)d\phi^2,
\end{eqnarray}
and then, the diagonal components of the induced metric tensor on the Beltrami surface are 
\begin{eqnarray}
    g_{uu}&=&\tanh ^2\left(\frac{u}{R}\right),\\
    g_{\phi\phi}&=&R^2 \sech^2\left(\frac{u}{R}\right).
\end{eqnarray}
The nonvanishing components of the Christoffel symbols are straightforwardly calculated and written as
\begin{eqnarray}
    \Gamma^{u}_{uu}&=&\frac{2 \csch\left(\frac{2 u}{R}\right)}{R},\\
    \Gamma^{u}_{\phi\phi}&=&2 R \csch\left(\frac{2 u}{R}\right),\\
    \Gamma^{\phi}_{u\phi}&=&\Gamma^{\phi}_{\phi u}=-\frac{\tanh \left(\frac{u}{R}\right)}{R}.
\end{eqnarray}
In order to properly take into account the effects of the geometry in the dynamics of a non-relativistic electron constrained to move on the Beltrami surface, we must compute the Da Costa potential, which is given by 
\begin{eqnarray}
   V_{dC}(u)=-\frac{\hbar^2}{8m^{*}R^2}\cosh ^2\left(\frac{u}{R}\right) \coth ^2\left(\frac{u}{R}\right).
\end{eqnarray}
It is worthwhile to mention that the da Costa potential $V_{dc}$ exhibits a parity-symmetrical potential well with respect to $u=0$. Such a well tends to trap the electron in a ring around $u=0$, as we can see in Fig.\ref{dacosta}. In the limit when $u\rightarrow 0$ the da Costa potential $V_{dC}\rightarrow -\infty$, such behaviour occurs due to the non differentiability of the surface in $u=0$. Hence, the da Costa potential tends to confine the particle around $u=0$ due to the edge in the surface. This behaviour is depicted in fig. (\ref{dacosta}) for three values of $R$, namely, $R=1$, $R=10$ and $R=100$. An important point that must be highlighted here is that the inclusion of the da Costa's potential is not in disagreement with the results found in \cite{Liu}. In \cite{Liu} the authors find that there is no geometric potential for a Dirac fermion on a two-dimensional curved surface of revolution. However, we are considering here a quadratic dispersion relation for the charge carriers, so that they are described by the Schr\"{o}dinger equation instead of Dirac equation. Hence, we must consider the da Costa's potential.

The axial symmetry leads to the periodic behavior of the wave function in the form
\begin{equation}
\label{axialsymmetry}
\Psi(u,\phi)=\psi(u)e^{i \ell\phi},
\end{equation}
where $\ell$ is the orbital quantum number. Substituting the Eq.\eqref{axialsymmetry} into Eq.\eqref{constantmassschrodinger}, the stationary Schr\"{o}dinger equation becomes
\begin{eqnarray}
\label{freenonhermitianequation}
-\frac{\hbar^2}{2m} \Lambda_1(u)\frac{d^2\phi(u)}{du^2}+\Lambda_2(u)\frac{d\phi(u)}{du}+\Lambda_3(u)\phi(u)=E\phi(u),
\end{eqnarray}
with $\Lambda_1(u)$, $\Lambda_2(u)$ and $\Lambda_3(u)$ given by:
\begin{eqnarray}
    \Lambda_1(u)&=&\coth ^2\left(\frac{u}{R}\right)\\
    \Lambda_2(u)&=&\frac{\hbar^2}{2 m R} \coth ^3\left(\frac{u}{R}\right)\\
    \Lambda_3(u)&=&\frac{\hbar^2}{8 m R^2} \cosh ^2\left(\frac{u}{R}\right) \left(4 l^2-\coth ^2\left(\frac{u}{R}\right)\right)
\end{eqnarray}

Note that the Eq.\ref{freenonhermitianequation} exhibits parity and time-reversal invariance, as a result of the Beltrami surface geometric symmetries. Nonetheless, the first order derivative terms render the Hamiltonian non-Hermitian with respect to the momentum $\hat{P}_u:= -i\hbar\partial_u$. The non-Hermiticity of the free electron Hamiltonian is not a problem, since the space-time reflection symmetry is preserved, the spectrum of the eigenvalues of the Hamiltonian is completely real \cite{Bender, Bender2}. Besides, there is an Hermitian equivalent Hamiltonian that can be achieved by a simple changing of variables. Considering the change in the wave function given by
\begin{equation}
    \phi(u)=\sqrt{\cosh \left(\frac{u}{R}\right)\tanh \left(\frac{u}{R}\right)}y(u)
\end{equation}
we are lead to the following one dimensional Hermitian Schr\"{o}dinger-like equation
\begin{equation}\label{modifiedschrodingerequation}
    \Lambda_1(u)\frac{d^2y(u)}{du^2} + V_{eff}(u)y(u)=Ey(u),
\end{equation}
where
\begin{eqnarray}
    V_{eff}=\frac{\hbar^2}{16 m R^2} \left[\cosh \left(\frac{2 u}{R}\right)+1\right] \left[4 l^2+3 \csch^4\left(\frac{u}{R}\right)-1\right].
\end{eqnarray}       
At this point we must highlight an important feature that becomes evident from the effective potential presented above, which is the fact that there is no break of chirality since the dependence on the orbital angular momentum is squared. 

\begin{figure}[h!]
    \centering
    \includegraphics[scale=0.8]{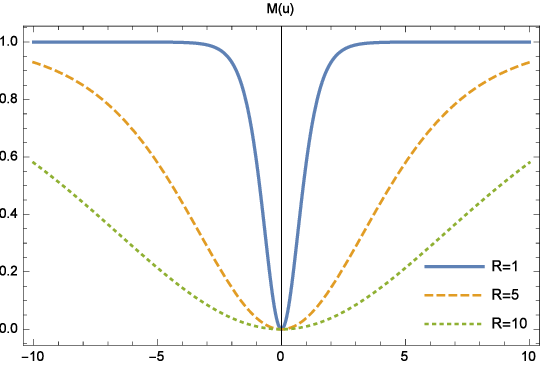}
    \caption{Behaviour of the position-dependent mass for three values of $R$, namely, $R=1$, $R=5$, $R=10$ and $m^*=1$.}
    \label{fig3}
\end{figure}

In order to write our dynamic equation in a more convenient way, let us perform the following transformation in the wave function, i.e., $y(u)=s(u)X(u)$, where $s(u)$ is given by $s(u)=e^{\frac{1}{2} \coth ^2\left(\frac{u}{R}\right)}$. By considering such transformation we are able to write the following equation
\begin{equation}\label{eq_pdm}
    -\frac{\hbar^2}{2}\frac{d}{du}\left[\frac{\Lambda_1(u)}{m^*}\frac{dX(u)}{du}\right]+\Bar{V}_{eff}(u)X(u)=EX(u),
\end{equation}
where $\Bar{V}_{eff}(u)$ is written as
\begin{equation}\label{effective_pot}
    \Bar{V}_{eff}(u)=V_{eff}(u)-\frac{\hbar^2}{2}\frac{\Lambda_1(u)}{m^*}\left[\frac{s''(u)}{s(u)}\right].
\end{equation}
Notice that the Schr\"{o}dinger equation as written in eq.(\ref{eq_pdm}), can be identified as describing a system whose mass is position-dependent \cite{pnbilayer, sinner, pdm1, pdm2, pdm3, pdm4, moraes}. In the present case, we can identify the position-dependent mass as being
\begin{equation}
    M(u)=\frac{m^*}{\Lambda_1(u)}.
\end{equation}
Such position-dependent mass exhibit an asymptotic behaviour so that when $u\rightarrow\pm\infty$ we have $M(u)\rightarrow m^*$. Also, when $u\rightarrow 0$, the effective mass reaches its minimum value. Such behaviour is a consequence of the singularity at $u=0$ and could be an indication that at $u=0$ the mean distance of the sites in the lattice structure of the material is greater in comparison to other regions of the surface. A greater mean distance between the sites of the lattice yields a smaller interaction with the electrons of the surface, which leads to a smaller effective mass. This is in agreement with more phenomenological considerations based on the Heisenberg uncertainty principle and the stretching of the manifold due to curvature, see f.e. \cite{dandoloff2, atanasov22}. The behaviour of the position-dependent mass is depicted in fig.(\ref{fig3}) for $m^*=1$ and three values of $R$, namely, $R=1$, $R=5$ and $R=10$.   

\subsection{Qualitative analysis}
Before obtaining the bound states and their respective spectra, let us discuss some qualitative features of the effective potential.

\begin{figure}[h!]
\begin{center}
\includegraphics[scale=0.8]{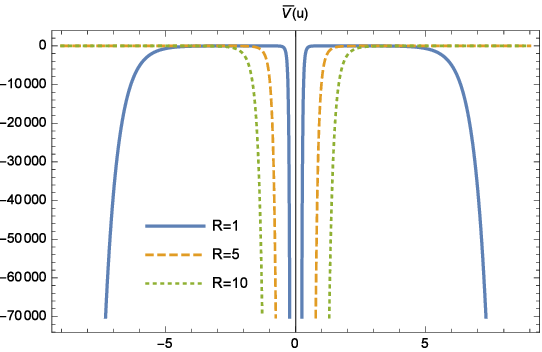}  
\caption{The effective potential for three values of $R$, namely, $R=1$, $R=5$ and $R=10$ and $\ell=0$, obtained by the eq.\ref{effective_pot}.}
\label{FIG2}
\end{center}
\end{figure}

In the fig.\ref{FIG2}, we show the effective potential for three values of $R$, namely, $R=1$, $R=5$ and $R=10$ and orbital angular momentum $\ell=0$, obtained by the eq.\ref{effective_pot}. We can see that the effective potential exhibit a behaviour qualitatively similar to the Da Costa potential presented in fig.(\ref{dacosta}). The same divergence is present at $u=0$ as an edge-like effect. An important feature that can be observed in fig.\ref{FIG2} is the fact that by increasing the value of $R$ the infinite well centered at $u=0$ becomes wider.

\begin{figure}[h!]
\begin{center}
\includegraphics[scale=0.8]{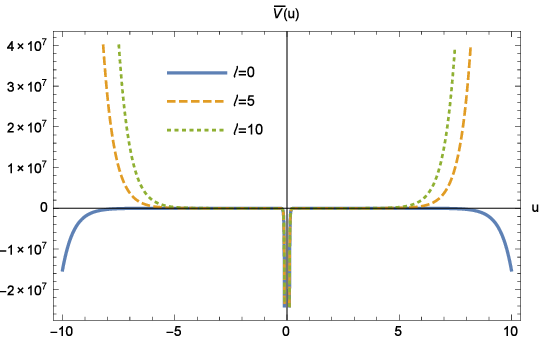}  
\caption{The effective potential for $R=1$ and three values of the orbital angular momentum, namely, $\ell=0$, $\ell=5$ and $\ell=10$, obtained by the eq.\ref{effective_pot}.}
\label{FIG3}
\end{center}
\end{figure}

The effective potential for $R=1$ and $\ell=0$, $\ell=5$ and $\ell=10$ is shown in the fig.\ref{FIG3}. In this figure, in addition to the geometric influence on the electron, there is the centrifugal potential emergent from the orbital angular momentum that significantly modifies the effective potential. When comparing the figures \ref{FIG2} and \ref{FIG3}, we observe that the centrifugal effect is suppressed for regions close to $u=0$. However, when we move away from $u=0$, the influence of the orbital angular momentum becomes more relevant and the effective potential starts to exhibit a prominent growth due to the dominance of the orbital angular momentum over the geometric effects.

%BOUND STATES--------------------------------------------------------------

\section{Bound states}

Let us investigate now the influence of both geometry and orbital angular momentum on the electronic states on the quantum Beltrami surface. For pedagogical purposes, let us study first the case with no orbital angular momentum, i.e., $\ell=0$ for three values of radius $R$, namely, $R=1$, $R=10$ and $R=20$. We have considered for all the plots $\hbar=1$ and $m^*=1$. 

\begin{figure*}[ht!]
    \centering
\begin{subfigure}{.33\textwidth}
  \centering
  \includegraphics[scale=0.6]{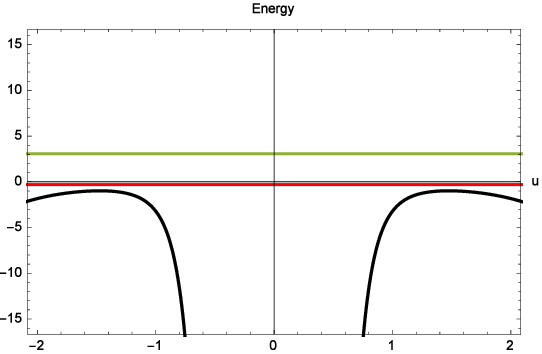}
  \caption{}
 
\end{subfigure}%
\begin{subfigure}{.33\textwidth}
  \centering
  \includegraphics[scale=0.6]{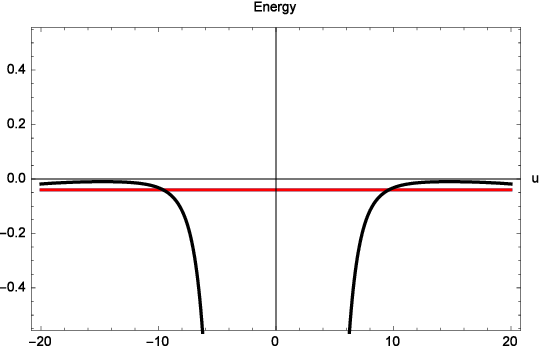}
  \caption{}

\end{subfigure}%
\begin{subfigure}{.33\textwidth}
  \centering
  \includegraphics[scale=0.6]{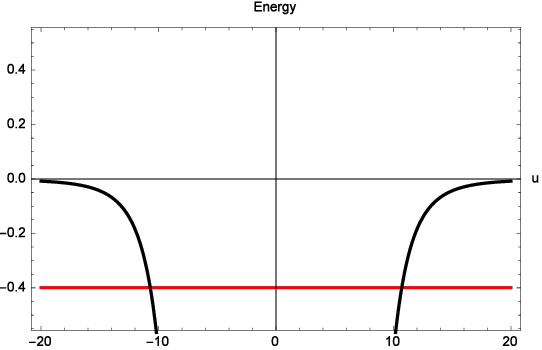}
  \caption{}

\end{subfigure}%
    \caption{Eigenenergies for the position-dependent Schr\"{o}dinger equation for three values of $R$, namely, $R=1$, $R=10$ and $R=20$. For this plot we have disregarded the orbital angular momentum. We have considered $\hbar=1$ and $m^*=1$.}
    \label{Fig11}
\end{figure*}

In fig. (\ref{Fig11}a) we are considering $R=1$ and no orbital angular momentum. For this configuration we can see that the first eigenvalues of the energy are not bound states, but propagating states. The infinite well centered at $u=0$ for $R=1$ is not wide enough to allow bound states. In a certain sense we can interpret this configuration as a reflectionless shaped potential.  

The fig. (\ref{Fig11}b) exhibit the first two states for $R=10$ and $\ell=0$. As we can see these bound states are practically degenerated and they are weakly confined at the well at $u=0$. But it is already possible to infer that by increasing the value of the radius $R$ more confined states become allowed. Finally, when we consider $R=20$ and $\ell=0$, as depicted in (\ref{Fig11}c), we can see a more confined bound state, confirming that as we increase the value of $R$ we allow the existence of more confined bound states.

\begin{figure*}[ht!]
    \centering
\begin{subfigure}{.33\textwidth}
  \centering
  \includegraphics[scale=0.6]{eigen_R1_L0.eps}
  \caption{}
 
\end{subfigure}%
\begin{subfigure}{.33\textwidth}
  \centering
  \includegraphics[scale=0.6]{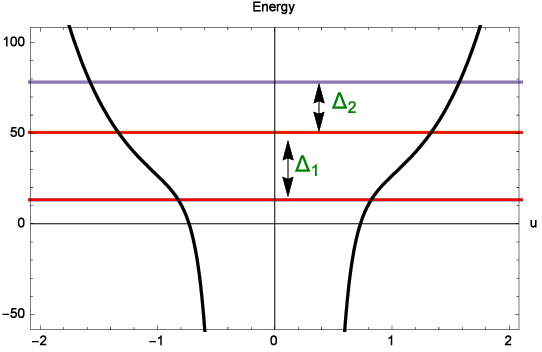}
  \caption{}

\end{subfigure}%
\begin{subfigure}{.33\textwidth}
  \centering
  \includegraphics[scale=0.6]{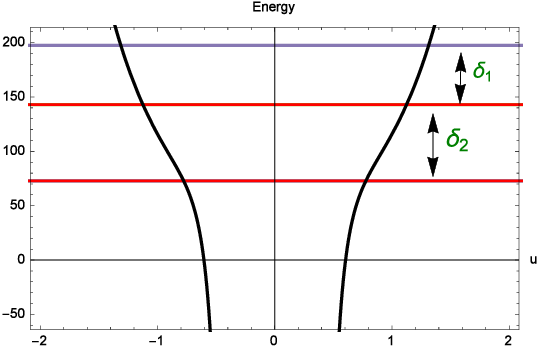}
  \caption{}

\end{subfigure}%
    \caption{Eigenenergies for the position-dependent Schr\"{o}dinger equation for $R=1$ and three values of orbital angular momentum $\ell$, namely, $\ell=0$, $\ell=5$ and $\ell=10$. For this plot we have considered $\hbar=1$ and $m^*=1$.}
    \label{Fig12}
\end{figure*}

In the figure (\ref{Fig12}) we have considered the effect of orbital angular momentum in the electronic states of the Beltrami surface. In order to do so, we have considered $R=1$ and three values of the orbital angular momentum, namely, $\ell=0$, $\ell=5$ and $\ell=10$. Also, for the plots we have considered $\hbar=1$ and $m^*=1$. In this plot we have added the case where $R=1$ and $\ell=0$, already discussed previously for the sake of comparison. 

In fig (\ref{Fig12}b) we have considered $R=1$ and orbital angular momentum $\ell=5$. Different from the case when $\ell=0$, as already discussed, the effect of the orbital angular momentum gives rise to a centrifugal potential that becomes dominant as we move away from $u=0$. The well formed due to the orbital angular momentum allows the existence of bound states. The first five bound states are depicted in (\ref{Fig12}b). It is important to highlight here that the first two bound states are degenerated (red line), as well as the third and fourth bound states. 

Finally, for $R=1$ and $\ell=10$ we have a similar behaviour in comparison to the case with $\ell=5$. The main difference between the two cases is the energy scale. For this case also, the well formed by the orbital angular momentum provides the possibility of bound states. The first two bound states (red line) are degenerated as well as the third and fourth bound state.

It is worthwhile to mention that the gaps in the eigenenergies $\Delta_1$ and $\Delta_2$, presented in (\ref{Fig12}b) are different, i.e., $\Delta_1\neq\Delta_2$. Such feature makes our system entirely different from a quantum harmonic oscillator, in which all the gaps between the eigenenergies are equal to $\hbar\omega$. This annharmonicity is an indication that the present system can fulfill the Di Vicenzo's criteria, which means that our system could be thought as a qubit suitable for encoding quantum information \cite{vicenzo}. Such a remarkable feature will be discussed in more details in a future paper.  

%FINAL REMARKS--------------------------------------------------------------

\section{Final Remarks}

In this paper, we investigate the influence of the geometry in the electronic states of a quantum Beltrami surface. We have considered an electron governed by the spinless stationary Schr\"{o}dinger equation constrained to move on the surface of a pseudosphere due to a confining potential from which the Da Costa potential emerges. We investigate the role played by the geometry and orbital angular momentum on the electronic states of the system. 

Initially we describe the geometry of the Beltrami surface and we show how the Da Costa potential, which is constructed by the mean and Gaussian curvatures, exhibit an infinite well at $u=0$ as some kind of edge effect. The Hamiltonian governing the electron dynamics was obtained directly from the geometry and it is non-Hermitian but $\mathcal{PT}$ symmetric. The invariance under parity reversion and time inversion allowed us to obtain an Hermitian equivalent Hamiltonian. 

The electron on the Beltrami surface behaves as if its mass depends on the position. The mass of the electron is smaller at $u=0$, also a consequence of the edge at $u=0$. The position dependency of the mass lead us to speculate that the effects of the lattice strain, which emerges as a consequence of the curvature, increase the mean distance of the sites of the lattice at $u=0$, which in its turn yield to a smaller mass at $u=0$.  

We have shown that the effective potential in the absence of orbital angular momentum exhibit a behaviour qualitatively similar to the Da Costa potential. The same divergence is present at $u = 0$ as an edge-like effect. An important feature that could be observed is the fact that by increasing the value of $R$ the infinite well centered at $u = 0$ becomes wider.

By considering the effects of orbital angular momentum we observe that the centrifugal effect is suppressed for regions close to u = 0. However, when we move away from $u = 0$, the influence of the orbital angular momentum becomes more relevant and the effective potential starts to exhibit a prominent growth due to the dominance of the orbital angular momentum over the geometric effects.

The bound state were numerically computed for the cases with and without orbital angular momentum. For the cases with no contribution of the orbital angular momentum, we could see that as we increase the value of the Beltrami radius $R$ we obtain more confined bound states. The consideration of a non-vanishing orbital angular momentum allows the existence of more confined bound states whose energy depends on the ratio between $R$ and $\ell$. An important feature that must be highlighted is the fact that there is an annharmonicity in the Hamiltonian of the system for a non-vanishing $\ell$. Such annharmonicity may lead us to the possibility of enconding a qubit in the energy states of the system. Such an important feature will be addressed in a future work.

\section{Acknowledgments*}

The author would like to thank J.E.G. Silva for valuable discussions, Alexandra Elbakyan and Sci-Hub, for removing all barriers in the way of science and the Funda\c{c}\~{a}o Cearense de Apoio ao Desenvolvimento Cient\'{i}fico e Tecnol\'{o}gico (FUNCAP) under the grant PRONEM PNE0112- 00085.01.00/16.

%BIBLIOGRAPHY--------------------------------------------------------------

\end{document}